\documentclass[structabstract]{aa}  
\usepackage{amssymb}
\usepackage{graphicx}
\usepackage{color}
\usepackage{txfonts}
\usepackage{subfigure}
\usepackage{natbib}
\newcommand{\nc}{\newcommand}
\nc{\lsun}{\ensuremath{\mathrm{L}_\odot}}
\nc{\msun}{\ensuremath{\mathrm{M}_\odot}}
\nc{\tex}{\ensuremath{\mathrm{T}_{\rm ex}}}
\nc{\cthree}{C$_3$}
\newcommand{\OI}{[O {\sc i}]}
\nc{\cthreehtwo}{$c$-C$_3$H$_2$}
\nc{\kms}{\mbox{km\,s$^{-1}$}}
\nc{\Kkms}{\mbox{K\,km\,s$^{-1}$}}
\nc\micron{\mbox{$\mu$m}}
\nc{\Trot}{$T_{\rm rot}$}%
\nc{\Ntot}{$N(C_3)$}%
\nc{\Tc}{$T_{\rm c}$}%
\nc{\Tdust}{$T_{\rm dust}$}%
\nc{\Tex}{$T_{\rm ex}$}%
\nc{\Tkin}{$T_{\rm kin}$}%
\nc{\Tmax}{$T_{\rm max}$}%
\nc{\cmcub}{\mbox{cm$^{-3}$}}
\nc{\cmsq}{\mbox{cm$^{-2}$}}

\def\ptsec{$''\mskip-7.6mu.\,$}
\newcommand{\HII}{H {\sc ii}}

\newcommand{\CII}{[C {\sc ii}]}
\newcommand{\CIIr}{C {\sc ii}}
\def\Msun{\,{\rm M$_{\odot}$}}

\def\Lsun{\,{\rm L$_{\odot}$}}

\nc{\thCO}{$^{13}$CO}
\begin{document}

   \title{CO outflows from young stars in the NGC\,2023 cluster}

   \subtitle{}

   \author{G. Sandell \inst{1} 
   \and B. Mookerjea \inst{2} 
   \and R. G\"usten \inst{3} 
    }

   \institute{ Institute for Astronomy, University of Hawaii at Manoa,  640 N. Aohoku Place, Hilo, HI 96720, USA \\
        \email{gsandell@hawaii.edu}
         \and
 Tata Institute of Fundamental Research, Homi Bhabha Road,
Mumbai 400005, India
                         \and
Max Planck Institut f\"ur Radioastronomie, Auf dem H\"ugel 69, 53121 Bonn, Germany     
             }

 \date{Last edit \today}
\abstract
{Young early-type HAeBe stars are still embedded in the molecular clouds in which they formed. They illuminate 
reflection nebulae, which shape the surrounding molecular cloud and may trigger star formation. They are therefore
ideal places to search for ongoing star formation activity.} 
{NGC\,2023 is  illuminated by the Herbig Be star HD\,37903. It is the most massive
member of a small young cluster with about 30 PMS stars,  several of which are Class
I objects that still heavily accrete. It might therefore be expected that they might drive
molecular outflows. We examined the whole region for outflows.}
{We analyzed previously published APEX data to search
for and characterize the outflows in the NGC\,2023 region. This is the first systematic search
for molecular outflows in this region. Since the
outflows were mapped in several CO transitions, we can determine 
their properties quite well.}
{We have discovered four molecular outflows in the vicinity of NGC\,2023, 
three of which are associated with Class I objects. MIR-63, a bright
mid-infrared and submillimeter  Class I source, is a binary with a separation
of  2\ptsec4 and drives two  bipolar outflows orthogonal to each other. The large southeast--northwest outflow excites the 
Herbig-Haro flow HH\,247. MIR-73, a Class I object, which is also a far-infrared source, drives 
a pole-on outflow.  MIR-62 is a Class II object with
strong infrared excess  and  a luminosity of 7 \Lsun. It is  not
detected in the far-infrared. The Class I sources have bolometric luminosities 
of about 20 \Lsun\ or lower, that is, they are all low-mass stars. One 
other far-infrared source, MIR-75, may have powered an outflow
in the past because it now illuminates an egg-shaped cavity.}
{The four outflows  are all powered by young stars  and are located in the immediate 
vicinity of NGC\,2023. They are at a similar evolutionary stage, which suggests that their
formation may have been triggered by the expanding \CIIr\ region. }
 \maketitle
\section{Introduction}

The reflection nebula  NGC\,2023 in the L\,1630 dark cloud is a  well-studied
reflection nebula that has been the source of several discoveries. The near-infrared
observations  by \citet{Sellgren84} showed that there must be
very small dust grains that are transiently heated by a single photon. These
are now generally known as polyaromatic hydrocarbons, or PAHs. Fullerenes, the
extremely stable C60 and C70 molecules, were also found in NGC\,2023
\citep{Sellgren11}. More importantly, however, it has served as a testbed for
developing and testing models of photodissociation regions (PDRs)
\citep{Black87, Draine96,Draine01,Kaufman06}.

NGC\,2023 is illuminated by the young B2 Ve star HD\,37903. This Herbig Be star
is the most massive member  of a young stellar cluster with $\sim$ 20 -- 30
pre-main-sequence (PMS) stars \citep{Sellgren83,DePoy90,Mookerjea09,Lopez-Garcia13}. The
reflection nebula is embedded in the L\,1630 dark cloud at a distance of $\sim$
400 pc. Most of the PMS stars in the cluster are  at the southern  and eastern
boundary of NGC\,2023, where the reflection nebula expands into the dense
surrounding molecular cloud, creating a hot PDR \citep{Sandell15}. Several of the
PMS stars are very young Class I sources \citep{Mookerjea09} that are heavily
accreting and likely to  drive molecular outflows. However, the outflows in the NGC\,2023 region 
have not been studied much so far. \citet{Malin87} found a chain of
HH objects that is now known as  HH\,247, in the southwestern boundary
of NGC\,2023.  They were not able to identify the exciting source of the HH
objects, although they speculated that  MIR-62  would be a likely
candidate because of its proximity to the HH objects. \citet{Mookerjea09}
suggested that the HH objects could be excited by the submillimeter/infrared source
MIR-63, but stated that this was not a clear identification.  We have used
the extensive APEX CO data sets from \citet{Sandell15} to search for and
characterize the molecular outflows from young stars in the NGC\,2023 cluster.
We find four molecular outflows and note that even HD\,37903 may drive a weak
molecular outflow.

\section{Observations}

\subsection{ APEX and SOFIA observations}

All the Atacama Pathfinder EXperiment, APEX\footnote{APEX, the Atacama
Pathfinder Experiment is a collaboration between the Max-Planck-Institut f\"ur
Radioastronomie, Onsala Space Observatory (OSO), and the European Southern
Observatory (ESO).}\citep{Gusten06}, and the SOFIA (Stratospheric Observatory
for Infrared Astronomy) CO(11--10) and \CII\  observations were discussed in
detail in \citet{Sandell15}. With APEX, we observed CO(3 --2), CO(4--3),
CO(6--5), CO(7--6), and $^{13}$CO(3--2), which are crucial for this paper.
 We additionally examined the SOFIA observations of \OI\  63 $\mu$m, which were
presented in \citet{Mookerjea23}. \CII\  is generally not an outflow tracer and
is not discussed further. Although the \OI\ fine structure line at 63
$\mu$m can be strongly enhanced in shocks and is often seen in jet-like molecular
outflows, it was not detected in any of the outflows either. In Table
~\ref{tab_lines} we list the basic information of the spectral lines that we
analyzed for molecular outflow activity. Further details of the
observations can be found in \citet{Sandell15} and \citet{Mookerjea23}. All the
spectral lines  discussed in this paper were calibrated in main-beam brightness
temperature. In order to compare the CO maps more easily, we smoothed the
CO(6--5) and CO(7--6) data by regridding them to the same spatial resolution 
as the CO(4--3) map, that is, 14\ptsec4. These smoothed maps are used in all the
figures, tables, and analyses presented in this paper.

\begin{table}[h]
\begin{center}
\caption{Observing setup. The quoted rms sensitivities are in T$_{mb}$ for a velocity resolution
of 0.5 \kms. $\theta_ {\rm FWHM}$  is the half-power beam width, and $\eta_{\rm mb}$ is the 
main-beam coupling efficiency. \label{tab_lines}} 
{\scriptsize
\begin{tabular}{llrrll}
\hline\hline

Receiver & Molecular transition &  Frequency   & $\theta_{\rm FWHM}$ & $\eta_{\rm mb}$ & rms\\
 & & [GHz] &  & & [K] \\
\hline
FLASH$^+$ & $^{13}$CO(3--2) & 330.587965 & 18.5 & 0.68 & 0.3 \\
FLASH$^+$ & CO(3--2) & 345.795990 & 17.7 & 0.68  & 0.2 \\
FLASH$^+$ & CO(4--3)   & 461.040768 & 13.3 & 0.58 & 0.4\\
CHAMP$^+$ & CO(6--5) & 691.473076 & 9.1 & 0.49 & 1.3 \\
CHAMP$^+$ & CO(7--6) & 806.651806 & 7.7 & 0.48 & 2.5 \\
GREAT L1  & CO(11--10) & 1267.014486 & 23.0 & 0.67 & 1.1 \\
upGREAT HFA & \OI\ 63 $\mu$m & 4744.777490 & 6.3 & 0.63 & 1.6\\
\hline
\hline
\end{tabular}}
\end{center}
\end{table}

\begin{table*}

\caption{Coordinates and UKIDSS magnitudes  of the outflow sources in NGC~2023. For saturated sources  (marked with $^a$), 
we quote 2MASS magnitudes. \label{tbl-NIR}}

{\small
\begin{center}
\begin{tabular}{lccccccc}
\hline
\hline
Source & $\alpha_{2000}$ & $\delta_{2000}$ & $Z$ & $Y$ & $J$ & $H$ & $K$ \\
 &[h m s] &[$^\circ$ \arcmin\ \arcsec ]   & [mag] & [mag]  &[mag] & [mag] & [mag] \\
\hline
MIR-62  &  5:41:36.390 &  -2:16:46.1 & 13.663 $\pm$ 0.002& 12.818 $\pm$  0.001 & 11.567 $\pm$  0.001&    10.049 $\pm$ 0.023$^a$ &   \phantom{1}8.83 $\pm$ 0.02$^a$  \\
MIR-63a &  5:41:37.178 &  -2:17:17.3& \ldots& \ldots& \ldots & 14.822 $\pm$ 0.010    &  11.24   $\pm$ 0.01\phantom{a} \\
MIR-63b &  5:41:37.028 &  -2:17:18.0 & \ldots& \ldots & 18.70 $\pm$ 0.18  &  14.127 $\pm$ 0.006 & 11.39 $\pm$ 0.01\phantom{a}   \\
MIR-73  &  5:41:44.601 &  -2:16:06.6 & 19.38 $\pm$ 0.13 & 17.92 $\pm$  0.04 & 15.832 $\pm$  0.012&     12.212 $\pm$ 0.001& 10.38 $\pm$ 0.03$^a$  \\
MIR-75  &  5:41:44.75 &  -2:15:54.8 & 17.39 $\pm$ 0.03 & 15.80 $\pm$  0.01 & 14.63 $\pm$  0.05 &       \phantom{1}11.44 $\pm$ 0.05$^a$ & \phantom{9}9.99 $\pm$ 0.03$^a$ \\
\hline
\hline
\end{tabular}
\end{center}
\noindent
$^a$ magnitude from 2MASS \\
}

\end{table*}

\subsection{UKIDDs images}

We retrieved the UKIRT Infrared Deep Sky Surveys (UKIDDS)\footnote{The UKIDSS
project is defined in \citet{Lawrence07}. UKIDDS uses the UKIRT Wide Field
Camera \citep[WFCAM;][]{Casali07} and a photometric system described in
\citet{Hewett06}.} calibrated images  as well as pipeline-reduced photometry  of
NGC\,2023 from the WFCAM science  archive\footnote{http://was.roe.ac.uk}. 
NGC\,2023 was observed as part of the Galactic Clusters Survey on 2005 October 14, 
in all UKIDDS filters. The near-infrared counterpart to MIR-63, a binary with a
separation of 2\ptsec4, did not have pipeline-reduced photometry. We
manually performed photometry using  GAIA\footnote{GAIA is a derivative of
the Skycat catalogue and image display tool, developed as part of the VLT
project at ESO. Skycat and GAIA are free software under the terms of the GNU
copyright.} and APT\footnote{http://www.aperturephotometry.org} on both
components using 2\arcsec\ apertures. The photometry for the outflow sources
discussed in this paper is given in Table~\ref{tbl-NIR}.

\subsection{{\it Herschel} PACS archive data}

NGC\,2023 was observed twice with the  {\it Herschel} Space
Observatory\footnote{{\it Herschel} was an ESA space observatory with science
instruments provided by European-led Principal Investigator consortia and with
important participation from NASA.}  in the guaranteed time key program, the
{\it Herschel} Gould Belt Survey (HGBS) \citep{Andre10,Bontemps10}. On
operational day (OD) 668 (2011 March 13) NGC\,2023 was observed  with fast scanning
(60\arcsec{}/sec) in parallel mode with PACS (70/160 $\mu$m; AOR 1342215984) and
SPIRE. We retrieved the PACS 70~$\mu$m level 2.5 JScanam image  from the {\it
Herschel} data archive. The field was additionally observed with medium scan
speed (20\arcsec{}/sec) and an orthogonal cross scan on OD 513 (2010 October 8) with
PACS (100/160 $\mu$m; AORs 1342206080 and 1342206081).  These data sets have a
much better image quality. Level 2.5 JScanam images were retrieved for both 100
$\mu$m  and 160 $\mu$m from the  {\it Herschel} data archive. All these data
sets were discussed in more detail by \citet{Stutz13}. Although PACS photometry
of NGC\,2023 sources was published by several groups
\citep{Furlan16,Konuves20,Fischer20}, the deduced flux
densities vary strongly, and we found one source, MIR-75, which is visible both at 70~$\mu$m and
100~$\mu$m, but has no entry in any of the above mentioned papers, nor is it found
in the  PACS point source catalog at  IRSA\footnote{https://irsa.ipac.caltech.edu}. 
We therefore performed photometry using a
combination of point-spread function fitting on background-subtracted images
using  GAIA and MIRIAD \citep{Sault95}. The PACS photometry is listed in
Table~\ref{tbl-PACS}.

\begin{table}
\begin{center}
\caption{Positions and flux densities of Herschel PACS  sources in
  the vicinity of NGC\,2023. Positions are derived from the 100 $\mu$m image. \label{tbl-PACS}}
{\scriptsize
\begin{tabular}{lllccc}
\hline
\hline
Name & $\alpha$(2000.0) & $\delta$(2000.0) & S(70 $\mu$m)& S(100 $\mu$m) & S(160 $\mu$m)  \\
  & [h m s] & [$^\circ$ \arcmin\ \arcsec ] & [Jy] & [Jy]  & [Jy]  \\
\hline
MIR-63     & 05:41:37.165 &  -2:17:16.3 & 36.5 $\pm$ 3.0& 35.8  $\pm$ 1.0  & 33.2 $\pm$ 3.8  \\
MIR-73     & 05:41:44.580 & -2:16:05.2 & 13.4 $\pm$  2.0      & 16.0 $\pm$ 0.7    & 18.2 $\pm$ 1.4  \\
MIR-75      & 05:41:44.760 & -2:15:55.3 & \phantom{1}4.3 $\pm$  0.4  & \phantom{1}5.4 $\pm$ 1.1  & \ldots  \\
\hline
\hline
\end{tabular}
}
\end{center}
\end{table}

\section{Molecular outflows}
\label{sect:outflows}

In the radio regime, CO is by far the best tracer of molecular outflows. CO is
one of the most abundant molecules after H$_2$, which has no dipole moment and
which therefore  has no observable transitions at radio wavelengths. CO is easily
excited in both diffuse and dense gas. The most  accurate mass estimates are
obtained by observing at least two rotational transitions and one line of
$^{13}$CO \citep{Bachiller92}. We have observed five rotational
transitions of $^{12}$CO and one $^{13}$CO transition. Molecular outflows 
can also be studied in a variety of other molecular lines, especially in strong shocks 
\citep[see e.g.][]{Bachiller97}, but even in these regions, the CO emission is
quite strong. For the purpose of this study, there was no need to observe any
other molecular tracer, especially since the nondetection of \OI\ 63\,$\mu$m, a
good shock tracer, demonstrates that there are no strong shocks in any of the
outflows.

\begin{figure*}
\begin{center}

\includegraphics[width=0.7\textwidth]{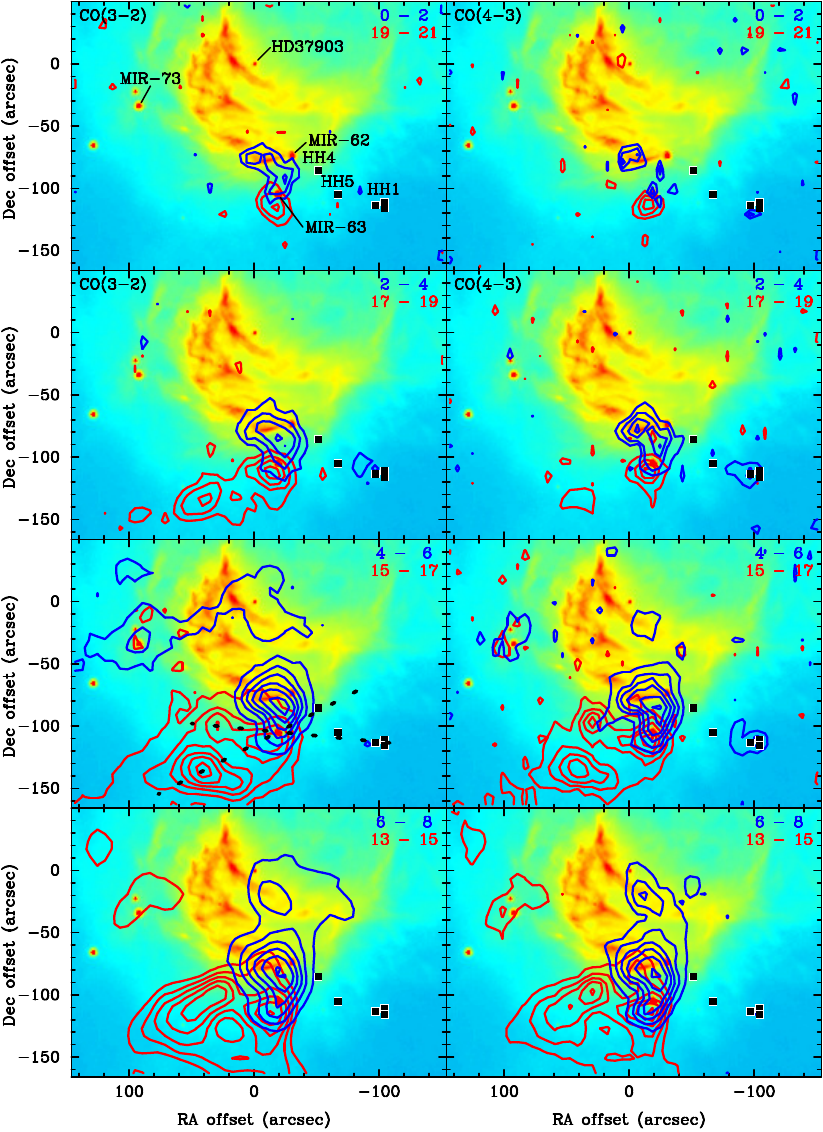}
\caption{Contours of high-velocity CO emission plotted in 2~\kms\ wide intervals
from a center velocity of 10.5~\kms\ and overlaid on the 8\,$\mu$m IRAC image in
color. HD\,37903 is at an offset 0\arcsec, 0\arcsec. The left panels show high-velocity 
emission in CO(3--2), and the right panels show this in
CO(4--3). The blue- and redshifted velocity ranges for the CO emission are
indicated in the top right corner in each plot. The positions of the young stars
MIR-63, MIR-62, and MIR-73 are labeled in the top left panel, and the
Herbig Haro objects are plotted with black squares.  The hatched lines in
the CO(3--2) panel for intermediate velocities are drawn to outline the bipolar
outflow from MIR-63.  The topmost  panels show three linearly spaced contours 
from 0.4~K to 1.1\,K in CO(3--2), and from 0.65\,K to 1.68\,K and  1.125\,K in red- and
blueshifted CO(4--3), respectively. At high velocities, $\pm$ 7.5 \kms, we plot 
four contours the from 0.6\,K for CO(3--2) and 0.7 K for CO(4--3), with the highest red 
and blue contours at  2.01\,K and 2.41\,K for CO(3--2) and 2.12\,K and 2.46\,K for CO(4--3). 
For low and intermediate velocities, we plot six contours 
starting at 0.8\,K for both CO(3--2) and CO(4--3) at intermediate velocities,
$\pm$ 5.5\,\kms, and at low velocities, $\pm$ 3.5\,\kms. The contours  start at 3\,K
and 2.5\,K to avoid contamination from the  molecular cloud. 
 For intermediate velocities, the highest contours are  at 6.1\,K and 6.0\,K and  4.9\,K and 6.1\,K 
 for CO(3--2) and CO(4--3), and for low velocities, they are
11.6\,K and 16.3\,K and 12.0\,K and 15.9\,K for CO(3--2) and CO(4--3), respectively.
\label{fig_outflows}}
\end{center}
\end{figure*}

\subsection{Overview}

In order to identify potential outflows more easily, we created channel maps from
the CO(3--2) and CO(4--3) images and overlaid the blue- and redshifted channel
maps on top of an 8\,$\mu$m IRAC image in color (see Figure~\ref{fig_outflows}).
This enabled us to see how the blue- and red-shifted high velocity gas are
related to each other and allowed us to identify potential bipolar outflows. In
Figure~\ref{fig_outflows}, we show  blue- and redshifted gas in the
velocity range 2 -- 8 \kms and 13 -- 21 \kms. We omitted the emission from 8 --
13  \kms, which is dominated by cloud emission. These velocity ranges capture
almost all of the high-velocity gas. The emission was averaged over 2~\kms\ wide
velocity intervals spaced symmetrically around a center velocity of 10.5\,\kms.
We only plot the part of the images  with  high-velocity gas. After we identified potential 
outflow sources, we examined the spectra in the
vicinity of that source and explored position-velocity diagrams, which is a good
way to obtain information about the velocity structure of the outflows.

The dominant outflow source is MIR-63. It is also a submillimeter
(MM\,3) and a far-infrared source. In the near-infrared, it is a heavily reddened
binary.  At low redshifted velocities, 13 -- 15 \kms,  a wide-angle
redshifted  outflow lobe is visible, centered on MIR-63 at a PA\footnote{Position Angle
(PA) is measured counterclockwise from north} of $\sim$ 130\degr, with a more
highly collimated blueshifted (6 -- 8\,\kms) lobe at a PA of $\sim$ -10\degr,
which appears to form a misaligned bipolar outflow. Some redshifted
emission also surrounds the Class I objects MIR-73 and MIR-75  at $\sim$
120\arcsec\ east of HD\,37903, where there are no young stars. At these
velocities, no gas is observed to be associated with the Herbig Haro flow
HH\,247\footnote{The Herbig Haro flow HH\,247 was originally called HH\,1, HH\,4
and HH\,5 by \citet{Malin87}}, where blueshifted CO
emission would be expected.

This picture changes at higher velocities. Faint blueshifted emission centered on the HH\,1 cluster 
is visible there, which was most likely
hidden in the relatively strong emission at near cloud velocities from the
molecular cloud surrounding NGC\,2023. Blue- and redshifted CO
emission is also visible toward MIR-73 and MIR-75  at intermediate velocities, $\pm$ 5.5 \kms, 
but not at higher velocities. A closer examination shows that MIR-73 powers this
low-velocity outflow. The blueshifted outflow lobe from MIR-63, which extended
all the way to HD\,37903 at low velocities, is far more compact at higher
velocities and appears to turn by 90\degr\ at the position of MIR-62 (see, e.g.,
CO(4--3) in the velocity range 4 -- 6\,\kms). At higher velocities, it seems to
align with a jet-like feature that is apparently associated with MIR-62, suggesting that
it is  a separate outflow driven by MIR-62.

At higher velocities, the redshifted outflow from MIR-63 appears to be more collimated,
with two prominent peaks east of the star. The northern peak, which dominated at
low velocities (13 -- 15\,\kms), is still strong at 15 -- 17\,\kms, but we also
see a southern peak that completely dominates the
emission at higher velocities. A line from the northern peak through MIR-63, that is, at a
PA of 84\degr, that is extended westward passes through HH\, suggesting
that the northern redshifted peak forms a bipolar outflow with HH\,1, which is part
of the blueshifted counterflow. Likewise, a line from the southern
peak  through MIR-63, that is, PA = 118 \degr,  that is extended westward
passes through HH\,4. The northern and southern redshifted
peaks appear to be part of a single outflow, whose blueshifted counterflow is the
Herbig Haro flow HH\,247. This is illustrated in Fig.~\ref{fig_outflows} in the
CO(3--2) panel for intermediate velocities, that is, the velocity range 4 -- 6
\kms\ in blue- and 15 -- 17\,\kms\ in redshifted CO. We outlined the outflow
by  the hatched black lines drawn through the two eastern redshifted CO peaks
and continuing through MIR-63 to the blueshifted side of the outflow, coinciding
with HH\,4 and HH\,1.

The blueshifted outflow lobe north of MIR-63 must be a 
separate outflow powered by MIR-63. It would therefore be expected to have a
redshifted counterflow to the south, which is indeed the case. At low
velocities, a tongue of emission protrudes from MIR-63 at a PA of
$\sim$170\degr, which agrees well with the northern blueshifted outflow, which
is at a PA of $\sim$ -10\degr. MIR-63 therefore drives two bipolar outflows, a
southeast-northwest outflow, and a north-south one. This is expected because 
we know that it is a binary.

Before we start to derive some basic parameters of these outflows, we
review what is known about the outflow sources and what we have learned 
from the near- and far-infrared photometry presented in this paper.

\subsection{Outflow sources}

\begin{figure}
\includegraphics[width=0.49\textwidth]{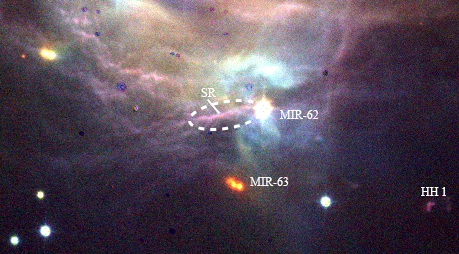}
\caption{UKIDDS color image (JHK)  of the MIR-63 and the MIR-62 region in the
southern part of NGC\,2023. Both stars are labeled. The very red binary system
in the middle of the image is MIR-63. The western component of the binary
coincides with the visible star. MIR-62 is $\sim$ 65\arcsec\ north of MIR-63.
We have labeled the bar  \citet{Sheffer11} called SR, the southern
ridge, which extends east of the star at a PA $\sim$ 110\degr. There is a
dark lane (cavity) on the opposite side.  We outlined the extent of the blue
outflow lobe centered on the southern ridge  with a hatched white ellipse. The
HH object that \citeauthor{Malin87} labeled HH\,1 is also labeled. In this image,
it appears as a faint extended  red feature. The size of the image is
183.6\arcsec\ $\times$ 101.6\arcsec. 
\label{fig_South_JHK}}
\end{figure}

\begin{figure}
\includegraphics[width=0.49\textwidth]{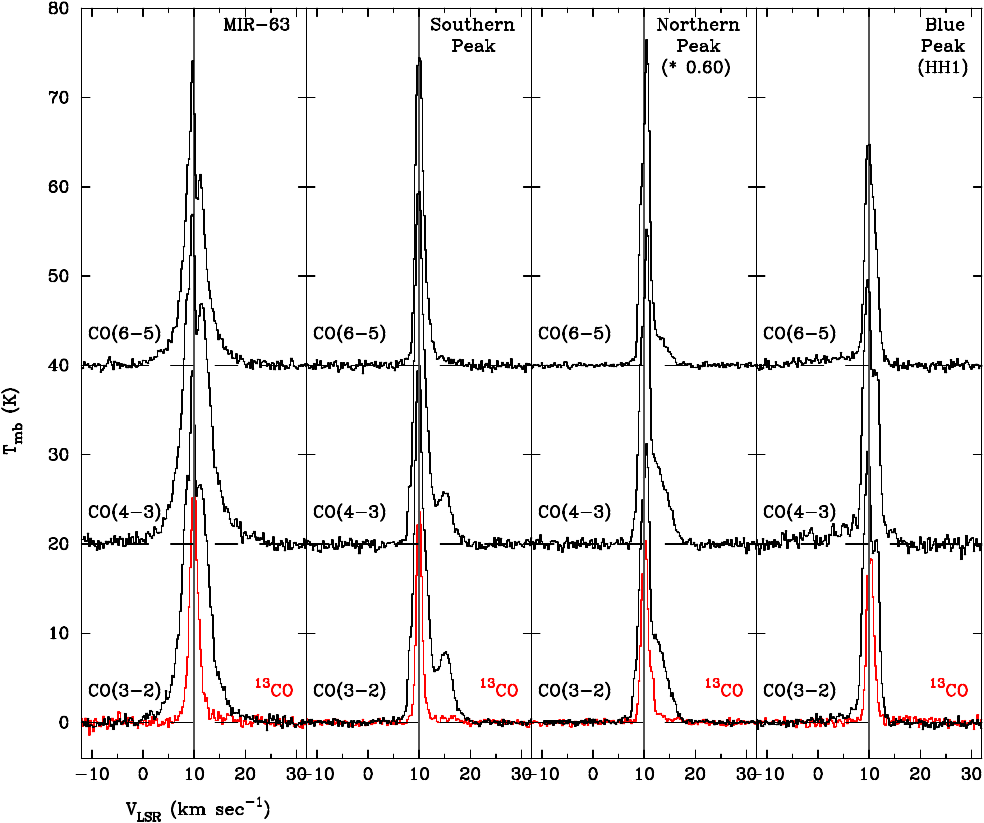}
\caption{Panel of CO spectra (CO(3--2),  $^{13}$CO(3--2), CO(4--3), and
CO(6--5)) toward MIR-63, the southern and northern peak in the red outflow,
and the  blueshifted emission peak near HH\,1. MIR-63 is at an offset 
-18\ptsec3, -104\ptsec7, the southern peak at 41\ptsec3, -133\ptsec0, the
northern peak at +30\arcsec,-97\arcsec, and HH\,1 at  -97\arcsec, -110\arcsec.
All offsets are relative to HD\,37903. The spectra toward the northern peak,
where the CO emission is very strong, are scaled by a factor of 0.6. The gray
vertical line marks the systemic velocity ($\sim$ 10.0 \kms). The spectra for
the different CO transitions are offset by 20~K. 
\label{fig_Dspectra}}
\end{figure}

\subsubsection{MIR-63}

MIR-63 is a bright mid-infrared source that  \citet{Mookerjea09} found to
coincide with the submillimeter source MM\,3. MM\,3 was first detected by
\citet{Wyrowski00}  at 3~mm with BIMA and also at  850 $\mu$m and  450
$\mu$m with  SCUBA.  At the time, there was no known optical or near-infrared
counterpart, and therefore \citeauthor{Wyrowski00}  assumed that it was most
likely a cold starless core. \citet{Mookerjea09} clearly identified it
with MIR-63, however, and they also associated Sellgren's star D with the same
mid-infrared source. They  classified MIR-63 as a Class I object and suggested
that it might be exciting the Herbig Haro flow HH\,247, originally named  HH\,5,
HH\,4, and HH\,1 by \citet{Malin87}. We now know that MIR-63 is a heavily
reddened binary in the near-infrared (Fig.~\ref{fig_South_JHK}). The eastern
component, MIR-63a, is associated with a nebulous arc and is only seen in the H
and K bands. This star dominates the emission in the mid- and
far-infrared. The western component, MIR-63b, which similarly is heavily
reddened, is also detected at J (Table~\ref{tbl-NIR}). MIR-63b
coincides with the optically visible star. It clearly contributes to the
emission at mid-infrared wavelengths and appears to have some emission even at
70 $\mu$m. In the far-infrared, MIR-63a is quite strong  and was detected by
PACS in all bands (Table~\ref{tbl-PACS}).

Although the source appeared to be extended in SCUBA observations, SCUBA did not have
the angular  resolution to resolve the source. Recently, MIR-63 was observed with
ALMA and the VLA as part of the  VANDAM survey of Orion protostars
\citep{Tobin20}. It was not detected by the VLA, but the continuum emission at
870~$\mu$m was resolved by ALMA into two sources, separated by 2\farcs4
\citep{Tobin20}. The brighter eastern component coincides with our far-infrared
source MIR-63a, while the fainter component coincides with MIR-63b. They marginally
resolve MIR-63a  as 0\farcs19 $\times$ 0\farcs12;  PA   =  3.1\degr, while b is
unresolved. MIR-63 was also included in the ALMA Survey of Orion Planck Galactic
Cold Clumps (ALMASOP), which observed 72 young dense cores  at 1.3~mm in three
different array configurations in continuum and molecular lines \citep{Dutta20}.
They found MIR-63 to be extended and identified four embedded sources in the
cloud core. Two of them are compact and coincide with MIR-63a and MIR-63b. These are
the same sources that  \citet{Tobin20} detected at  870 $\mu$m. The other two
were extended, $\sim$1\arcsec\ to $>$3\arcsec, and appear to be starless cores.
They are within 500 au of the two young heavily accreting Class I protostars,
both of which drive outflows. Their temperatures are therefore likely to be
around 25 to 30 K.  A recent paper \citet{Luo23} reanalyzed the
\citeauthor{Dutta20} ALMA data for  MIR-63 and claimed that it drives a
complicated molecular outflow that is viewed pole-on.  Their analysis  of  the
starless cores showed that they are transient structures for any reasonable
assumption of gas temperature.  As we already saw in the previous section, MIR-63
drives two molecular outflows, a large southeast-northwest (SE-NW) outflow, and
a more compact north-south (N-S) outflow. Neither of them are close to pole-on.
The SE-NW component has an inclination of $\sim$ 70\degr (see below). The N-S outflow
probably has a similar inclination because the red- and blueshifted outflow
lobes are spatially well separated.

We derive a bolometric luminosity of  33.0 $\pm$ 1.2 \Lsun\ using the photometry
in Tables~\ref{tbl-NIR} \& \ref{tbl-PACS} supplemented with Spitzer IRAC and
MIPS photometry from \citet{Mookerjea09}. Additionally, we used the tabulated
Spitzer IRS photometry points from \citet{Furlan16}. The latter showed that MM\,3,
or MIR-63a, has a fairly strong silicate absorption feature, indicating
that the disk is seen somewhat edge-on. The bolometric temperature is around
200\,K for an assumed foreground extinction of 10 mag. \citet{Furlan16} derived
a bolometric luminosity of 28.2 \Lsun. This is slightly lower than what we
derive, most likely because of differences in the IRAC/MIPS flux densities.  The
modeling by \citet{Furlan16} suggests that the outflow has an inclination of
70\degr, which sounds very plausible.

In the overview section, we interpreted the outflow emission from MIR-63 seen in
Fig.~\ref{fig_outflows} as being composed of two outflows. The large
redshifted outflow lobe east of the source is one bipolar outflow, and the
blueshifted lobe is faint but still visible in the Herbig-Haro flow HH\,247.
This outflow is most likely driven by MIR-63a, which was resolved with ALMA by
\citet{Tobin20} to have a roughly north-south oriented disk (PA $\sim$3\degr{}).
The second outflow goes roughly north-south, with the blueshifted outflow lobe
to the north and the redshifted lobe to south, partly overlapping with the large
eastern redshifted outflow lobe. The position-velocity diagram that goes
through the northeastern redshifted peak and the HH objects to the west, that is, PA
 84\degr, (Fig.~\ref{fig_DNpvplot}), shows that to the east, the outflow lobe
terminates at $\sim$ 80\arcsec\ from MIR-63, while the blueshifted outflow lobe
extends to $\sim$100\arcsec. The position-velocity diagram through the
southeastern redshifted peak (Fig.~\ref{fig_DSpvplot}), that is, at a PA of
118\degr, shows that the redshifted emission is still quite strong at the edge
of our CO maps, at 100\arcsec\ from MIR-63, indicating that we do not fully
cover the southeastern outflow lobe. In  the following, we assume that the
redshifted outflow has a total extent of 110\arcsec. There  is some
blueshifted emission to the west, but except for the high-velocity gas
associated with HH\,1 the blueshifted outflow is largely invisible.

At the position of MIR-63, CO(3--2),  CO(4--3), and CO(6--5) show strong
blue- and redshifted high-velocity wings extending from about  -5 to +28
km~s$^{-1}$ (Fig. \ref{fig_Dspectra}). The systemic velocity, determined from
$^{13}$CO(3--2), is $\sim$ 10 km~s$^{-1}$.  Only faint high-velocity emission at
near cloud velocities is seen in  CO(7--6), and  CO(11--10) shows no sign of 
high-velocity emission. The $^{13}$CO(3--2) emission shows clear redshifted high
velocity wings not only on MM\,3, but also on the northern and southern
redshifted peak  (Fig.\,\ref{fig_Dspectra}), indicating that the redshifted
high-velocity gas is somewhat optically thick. At the blueshifted high-velocity
peak, which roughly coincides with the HH\,1 cluster, the blueshifted emission
is stronger in CO(6--5) and CO(4--3) than in CO(3--2), indicating that the
blueshifted high-velocity gas in this region is warmer than in the rest of the
outflow. Even CO(7--6) shows a blueshifted emission wing extending to a velocity
of -8 \kms. This is  the only region where high-velocity emission is seen in
CO(7--6).

The N-S outflow, PA $\sim$  -10\degr\ from MIR-63, is less extended. The
blueshifted outflow lobe, which extends to $\sim$ 60\arcsec\ from MIR-63, shows
high-velocity gas to $\sim$ -4 \kms\ (total velocity 14 \kms{}), while the
redshifted gas extends to +28 \kms (total velocity 18 \kms{}; see
Fig.~\ref{fig_NSpvplot}). However, since the redshifted outflow lobe overlaps
with the redshifted southeastern  outflow lobe, some of the emission could also
be from the  southeastern outflow. There is still high-velocity gas at the edge
of our map, indicating that the outflow extends farther than 45\arcsec\ from
MIR-63. In our analysis, we assumed that it terminates 55\arcsec\ from MIR-63.  As
we already mentioned, the N-S outflow is most likely driven by MIR-63b, the
optically visible star in the MIR-63 core.

\begin{figure}
\includegraphics[width=0.49\textwidth]{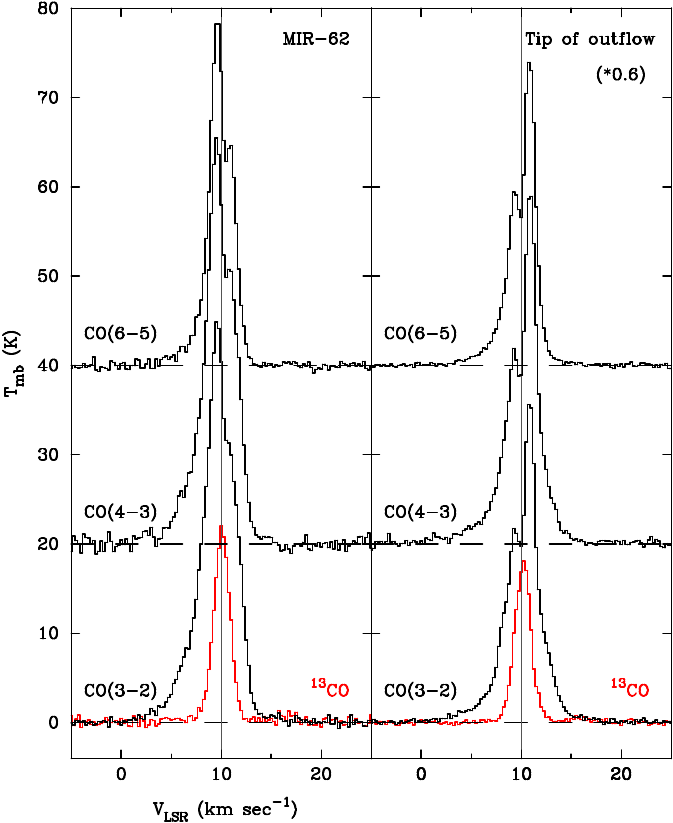}
\caption{CO spectra toward MIR-62 at an offset -30\ptsec0, -74\ptsec6, and at the
tip of the blue outflow lobe at an offset +4\ptsec0, -81\ptsec1. The spectra at the
tip of the outflow are scaled by a factor of 0.6,  as indicated in the figure.
The CO transitions are offset by 20 K for clarity.  The $^{13}$CO(3--2) spectra 
are plotted in red. The vertical gray line marks the systemic velocity ($\sim$
10 \kms{}) on MIR-62.
\label{fig_Cspectra}}
\end{figure}

\subsubsection{MIR-62}

MIR-62 (Sellgren's star C; Haro\,5-49) is also a binary with a separation of
$\sim$ 0\ptsec35 \citep{Sheffer11,Kounkel16}. It has strong infrared excess and was
classified as a Class I/II object by \citet{Mookerjea09}. \citet{Megeath12}
classified it as a Class II source and estimated a foreground extinction of
$\sim$ 5 mag. It was not detected by PACS at 70 $\mu$m, and it is not a
submillimeter source. \citet{Scarrott89} found that it illuminates a small
bipolar reflection nebula oriented north-south (see the JHK image in
Fig.~\ref{fig_South_JHK}.) This would suggest that it might drive a north-south
outflow, but as we described above, it drives an east-west outflow.  However, if the
outflow is largely in the plane of the sky, then light would be reflected
from the cavity walls of the east-west outflow. This is most likely the case.
\citet{Kounkel19} found it to have a spectral type of M3. We determine a
bolometric luminosity of 6.7 $\pm$ 0.1 \Lsun\ and a bolometric temperature of
$\sim$1900 K for an assumed foreground extinction of 5 mag.

Fig.~\ref{fig_outflows} showed that the star appears to power a blueshifted
outflow lobe to the east. This  outflow lobe  is centered on a bright ridge
pointing toward MIR-62 (see Fig. ~\ref{fig_South_JHK}).  \citet{Sheffer11} argued
that this is a PDR feature and named it the southern ridge (SR) because studies
of vibrationally  excited H$_2$ emission \citep{McCartney99,Burton90} and
purely rotational H$_2$ lines \citep{Sheffer11} showed that these lines are
dominated by PDR emission in the southern ridge and are not due to shock excitation.
It is a strange coincidence that the southern ridge is so well aligned with the
blueshifted outflow lobe if it is not associated with the outflow. The eastern
end of the ridge is  35\arcsec\ from MIR-62, while we estimate the length of the
outflow to be  $\sim$ 45\arcsec\ (0.087 pc) from the position-velocity diagram
in Fig.~\ref{fig_Cpvplot}.

The outflow velocities  are rather modest. At the tip of the outflow,
they extend from $\sim$ -2  to +16 \kms\ (Fig.\,\ref{fig_Cspectra}). There
is some redshifted emission at the position of MIR-62, but there is no sign of
a redshifted counterflow (Fig.\,\ref{fig_Cpvplot}). The UKIDDS JHK color image
in Fig.~\ref{fig_South_JHK} shows a narrow dark structure west of the star, 
which is approximately aligned with the blueshifted
outflow lobe, and which opens up to a faintly limb-brightened cavity
farther northwest (like a wine glass). This is most likely created by the
counterflow from MIR-62. It is seen even more clearly in the JHK 
image of \citet{Sheffer11}, their Fig.\,1.

 \begin{table*}
 \begin{center}
\caption{Physical parameters of the outflows. Dynamical timescales and
mass-loss rates are computed using mass weighted velocities.
All derived parameters are corrected for inclination. The formulae for the inclination 
correction factors are given in \citet{Bachiller90}.  \label{tbl-outflows}}
\begin{tabular}{lccccccc}
\noalign{\smallskip}
\hline\hline\noalign{\medskip}
\multicolumn{1}{l}{Parameter}&
\multicolumn{2}{c}{MIR-63a}&
\multicolumn{2}{c}{MIR-63b}&
\multicolumn{1}{c}{MIR-62}&
\multicolumn{2}{c}{MIR-73}\\
                  &  red      & blue & red & blue & blue & red & blue\\
\noalign{\medskip}
\hline\noalign{\medskip}
Maximum outflow velocity (\kms)   &  18        & 15         &  18 & 14 & 12 & 5.6  & 5.4\\
Inclination (\degr)                           & 70$^a$ & 70$^a$ & 70  & 70 & 70 & 17.6  & 2.4\\
Temperature (K)                             &  30        & 75        & 30  & 30 & 40 & 30   & 40\\
Dynamical time scale (10$^3$ yr)  &   5.6    & n.a.         & 3.1 & 4.3 &  3.2 & 14.4 & 12.3\\
Mass (10$^{-3}$ \Msun)                 & 130     &  2.4$^b$ & 52  & 48 & 45 &  2.4 & 1.5\\
Momentum (10$^{-2}$ \Msun\ km s$^{-1}$) & 160 & n.a. & 63 & 47 & 44 & 0.99 & 0.67\\
Force (10$^{-5}$ \Msun\ km s$^{-1}$ yr$^{-1})$ & 10.0 & n.a. & 7.4 & 3.8  & 4.8 & 0.055 & 0.073 \\
Energy (10$^{36}$ J)                    & 50.4 & n.a. & 8.6 & 5.0 & 1.7 & 0.014 &0.031\\
Mass loss rate (10$^{-6}$ \Msun\ yr$^{-1}$)  & 23 & n.a.& 17 &  11 &  14 & 0.17 & 0.12\\
\noalign{\medskip}
\hline
\end{tabular}

$^a$ Outflow inclination from \citet{Furlan16}; 
$^b$ Mass of HH\,1 condensation only \\
\end{center}
\end{table*}

\subsubsection{MIR-73}

\begin{figure}
\includegraphics[width=0.49\textwidth]{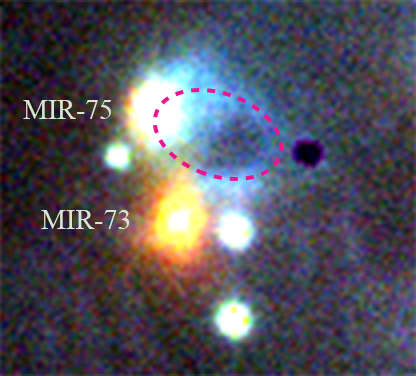}
\caption{UKIDDS color image (JHK)  of the MIR-73 and MIR-75 region east of
HD\,37903. Both are Class I sources with strong infrared excess, and both were
detected by PACS. The stars are labeled. MIR-73, the very red source in the
middle of the image, drives a compact  almost pole-on outflow. MIR-75 does
not appear to drive a molecular outflow, but sits at the apex of a bubble of
emission, which we outlined with a dashed red ellipse. The hole west of
it is an image artifact. The size of the image is 41\ptsec6 $\times$ 37\ptsec6.
\label{fig-NE_JHK}}
\end{figure}

MIR--73, on the eastern side of the reflection  nebula (Fig.~\ref{fig-NE_JHK}),
was classified as a Class I object by \citet{Mookerjea09}. It has strong near-infrared
excess and is relatively bright in the far-infrared. It was easily detected by PACS at
all bands (Table~\ref{tbl-PACS}). It was not detected at 850 $\mu$m by SCUBA
\citep{Johnstone06}, but \citet{Furlan16} reported a marginal detection at 350 
$\mu$m. \citet{Furlan16} derived a bolometric luminosity of 17.8 \Lsun, and a
bolometric temperature of 277 K for an assumed foreground extinction of 16 mag.
This agrees reasonably well with what we derive, a bolometric luminosity of
20.9 $\pm$ 0.1 \Lsun\ and a bolometric temperature of about 260 K  for a
foreground extinction of 10 mag.

\begin{figure}
\centering
\includegraphics[width=0.25\textwidth]{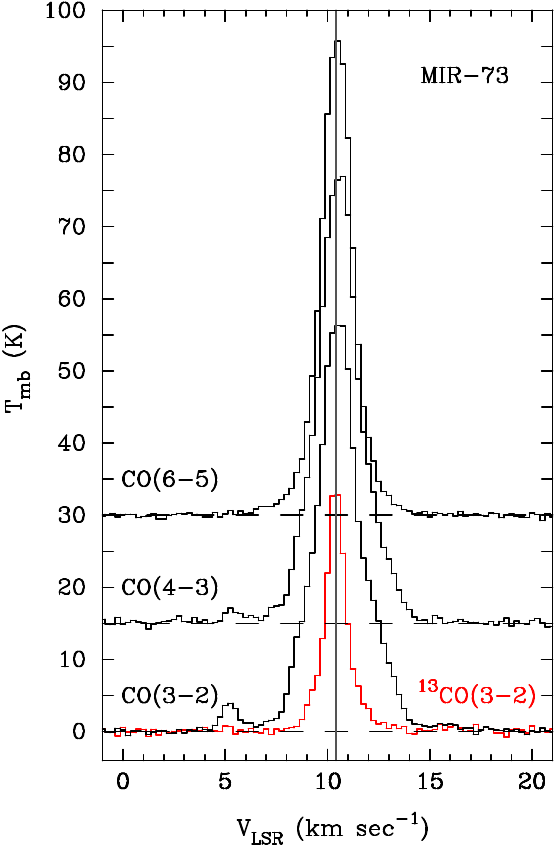}
\caption{CO spectra toward MIR-73. The left panel shows CO(3--2),
and the right panel shows CO(4--3). The $^{13}$CO(3--2) spectrum  is also
plotted in red in the left panel. The vertical gray line marks the
systemic velocity, $\sim$ 10.4 \kms. The narrow emission feature at 5.2
\kms\ is due to an IR dark cloud filament and
not related to the outflow from MIR-73. 
\label{fig_MIR73spectra}}
\end{figure}

\begin{figure}
\includegraphics[width=0.49\textwidth]{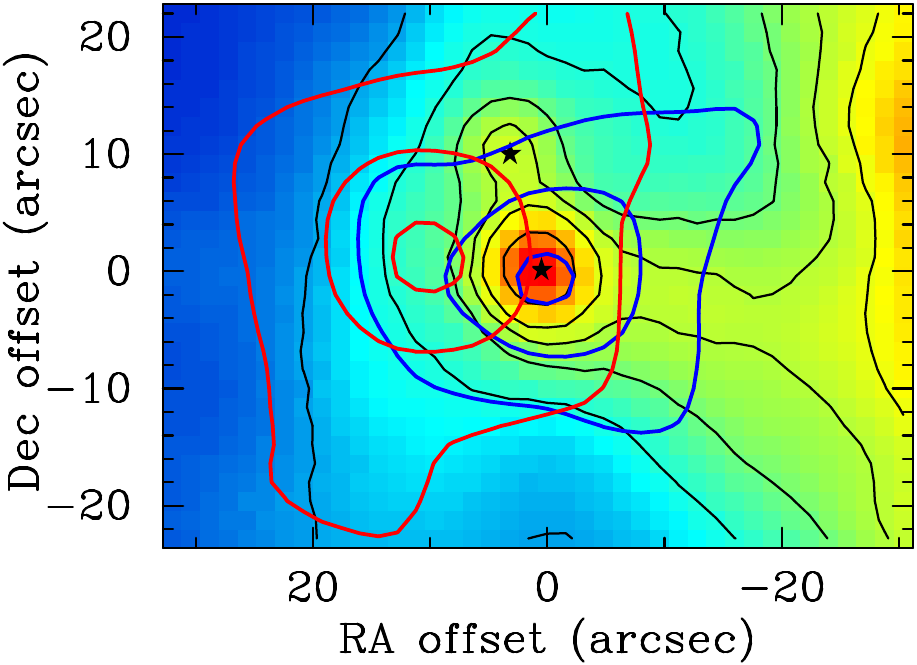}
\caption{High-velocity emission of CO(6--5) overlaid on a 100 $\mu$m PACS
image in color. The blueshifted CO is integrated from  +6 to + 8 \kms, and the
redshifted CO is integrated from 13--15 \kms. This CO(6--5) map was created with the same 
angular resolution as  CO(4--3) or 14\ptsec4. MIR-73 is placed at an offset
0\arcsec, 0\arcsec\ , and MIR-75 is $\sim$ 12\arcsec\ north of it. Both are
marked by black stars.
\label{fig-mir73outflow}}
\end{figure}

The systemic velocity of the molecular gas surrounding MIR-73 is $\sim$
10.4\,\kms (see Fig.~\ref{fig_MIR73spectra}). The blue- and redshifted emission
lobes partly overlap. The separation between the red- and
blueshifted peak is 10\farcs1 , suggesting that the outflow is seen almost pole-on 
(see Fig.~\ref{fig-mir73outflow}). The outflow models by \citet{Cabrit86} make an
inclination angle of 20\degr\ appear about right. The blueshifted peak is almost centered on
the star, offset  by only  1\farcs2 to the west,  while the redshifted  outflow lobe is off by 8\farcs9 to the east.
We therefore assign an inclination of  2.4\degr\ to the blueshifted outflow and leave 17.6\degr\ for the redshifted outflow.
The outflow velocities are very modest, $\sim$ 5 -- 6  \kms\ for both
outflow lobes (see Fig.~\ref{fig_MIR73spectra}).

\subsubsection{MIR-75}

MIR-75 was classified as a Class I object by \citet{Mookerjea09}.
We see no high-velocity CO toward the star, but it sits at the apex of a
limb-brightened cavity on the western side of the star (see
Fig.~\ref{fig-NE_JHK}). This egg-shaped cavity is seen in all UKIDDS bands,
suggesting that it is illuminated by reflected light from MIR-75. It
appears likely that a past outflow may have created the cavity, although there
is currently no evidence for outflow activity. MIR-75 is also detected with
PACS at 70 and 100 $\mu$m and is most likely present at 160 $\mu$m as well, but
the emission from the star cannot be separated from the strong emission from the
surrounding cloud. We derive a bolometric luminosity of 3.6 $\pm$ 0.2 \Lsun\ and
a bolometric temperature of 825 $\pm$  9 K, assuming that the foreground
extinction is 10 mag, which is similar to the foreground extinction toward MIR-73.

There is no sign of a cavity on the eastern side, possibly because it is
obscured by the IR dark cloud filament just east of the star. Another
possibility is that MIR-75 is near the edge of the NGC\,2023 cloud, so that the
outflow expanded more freely into the low-density cloud envelope.

\subsubsection{HD\,37903}

\citet{Wyrowski00} found a weak extended ($\sim$ 40\arcsec{}) continuum source
at 8\,GHz centered  at $\sim$ -10\arcsec, -12\arcsec\ relative to HD\,37903.
They interpreted this continuum emission as free-free emission from a density
enhancement  ionized by the Lyman continuum flux from HD\,37903. However,  PDR
tracers such as \CII\ and \OI\ do not show any enhancement from this continuum
source \citep{Sandell15,Mookerjea23}. As we showed in
Fig.~\ref{fig_outflows}, a blueshifted emission feature lies southwest of the
star extending to $\sim$ 4\,\kms, that is, $\sim$ 6.5\,\kms\ from the systemic
velocity. This blueshifted emission peaks roughly at -16\arcsec, -24\arcsec\
from HD\,37903 and has a size of $\sim$ 50\arcsec. It is in the same direction
as the low brightness continuum emission, suggesting  that they may be related.
Since we see it in  CO(3--2), CO(4--3) and CO(6--5), we can estimate the
temperature assuming LTE and optically thin emission (see
Section~\ref{sect:properties}). This gives us an excitation temperature of
$\sim$ 30 K. If the blueshifted CO emission were associated with the ionized
emission feature, the gas would be expected to be warmer. The cause for this 
blueshifted CO emission is unclear. Maybe it is just being
pushed by the stellar wind from HD\,37903. The total mass of the mass of the
blueshifted gas$\sim$ 0.008 \Msun, which is on the same order as mass in the
MIR-73 outflow (see Table~\ref{tbl-outflows}).

\subsection{Properties of the outflows}
\label{sect:properties}

Since we detect all outflows in CO(3--2), CO(4--3), and CO(6--5) and in most
cases do not see any high-velocity gas in CO(7--6),  we know that the outflowing
gas must be warmer than 20\,K, but probably not much warmer than 40\,K.  There is
one exception, however. The blueshifted high-velocity gas associated with the
HH\,1 cluster is detected even in CO(7--6), and here, we derive a temperature of
$\sim$ 75\,K. The redshifted counterflow is much colder. To estimate the
column densities, we integrated over the outflows  in 2\,\kms\ wide velocity
intervals from -1 to 22 \kms, except for the velocity range 8 -- 13 \kms, which is
dominated by strong molecular cloud emission. For regions in which the outflows
overlap, we assumed equal contribution from each outflow. We repeated this for
$^{13}$CO(3--2), although we only detected $^{13}$CO(3--2) line wings in the
near cloud velocity ranges, that is, from 6 -- 8 \kms\ in blueshifted emission and
13 -- 15 \kms\ in redshifted emission. We can now solve for the optical depth
from the ratio of CO/ $^{13}$CO by assuming that both have the same excitation
temperature and that we know the isotope ratio [C/$^{13}$C], which we assume is
70  (see, e.g., \citet{Sandell15}). In this way, we find that all outflows except
MIR-73 are somewhat optically thick at low velocities, with optical depths in
the range 2.5 -- 6 and a median of around 3.5. For higher velocities, our upper
limits suggest that the emission is largely optically thin. We can now determine
the  excitation temperatures and CO column densities by assuming that CO is
in LTE. We can now use the formulae for calculating column densities given by
\citet{Mangum15} for optically thin emission and assuming the excitation
temperature is the same for all velocities over each outflow lobe. To do this, we
calculated the column density for a grid of temperatures from 25 -- 50 K for each
CO transition, and then we determined the resulting temperature within the errors
with the same column density for each CO transition. For the initial analysis, we
gave the near cloud velocities lower weight because we know that they are
optically thick. After we constrained an excitation temperature (Table~\ref{tbl-outflows}), 
we calculated
optical depth corrections for CO(4--3)  and  CO(6--5) to obtain optical depth-corrected 
column densities. For each velocity bin, we took the average of the
total column densities derived for each transition. The masses given in 
Table~\ref{tbl-outflows} assume that the CO/H$_2$ ratio
is 10$^{-4}$ and that the total atomic weight per hydrogen atom is 1.4
\citep{Kauffmann08}.

We calculated the outflow parameters largely following the procedures outlined by
\citet{Lada85}, \citet{Bachiller90}, and \citet{Choi93}. To derive the mass,
momentum, and and energy, we integrated over the same velocity intervals as  we
use in Fig.~\ref{fig_outflows}. All outflow parameters were corrected for
inclination using the formulas given in \citet{Bachiller90} (see also
\citet{Cabrit92}). To determine the dynamical timescales, we used the mass-weighted
outflow velocity rather than the maximum velocity. The difference compared to using
the maximum outflow velocity for our outflow sources is smaller than 50\%. The outflow
parameters are given in Table~\ref{tbl-outflows}.

In the review by \citet{Bachiller92}, they found that a comparison of different
methods for determining the accuracy of the outflow parameters by Cabrit and her
collaborators \citep{Cabrit86,Cabrit88,Cabrit92} indicated that the most
accurate methods estimate the mass within a factor of 2, the momentum within a
factor of 10, and the force within a factor of 20. Even with the extensive CO
data set we have, we do not appear to do much better. Our mass estimates are
uncertain because of the overlap of outflows and because we average the emission
over the whole outflow lobe. This results in large uncertainties in the optical
depth correction for gas at low velocities. Furthermore, we are somewhat
overcautious in setting the boundary between outflow emission and the ambient
cloud, which may contain the bulk of the outflowing gas, resulting in a
uncertainty of about a factor of 3. For other outflow parameters, the
uncertainty in inclination is the dominant error source. We assumed an
inclination angle of 70\degr\ for the two MIR-63 outflows as well as for MIR-62.
However, the inclination angle for these sources could be as  high as 80\degr\ or as low as
50\degr, resulting in uncertainty factors of 4 for momentum, 14 for energy, and
18 for the outflow force, just from the uncertainty in inclination. For MIR-73,
the almost pole-on outflow, the errors in momentum and energy due to inclination
are small, while the dynamical timescale  and force  is uncertain by a factor of
10 or more. 

\section{Discussion}

All the molecular outflows discovered in the NGC\,2023 region are in the
immediate periphery of the  \CIIr\ region powered by HD\,37903. Furthermore, they
all lie in the eastern and southern part of nebula, where the \CIIr\
region expands into the dense surrounding molecular cloud. This suggests
that their formation may have been triggered by the  \CIIr\ region. 
\citet{Elmegreen77} introduced the concept of triggered star formation when they
investigated the formation of  OB associations, which appear to follow a
sequential  star-forming mechanism. In this scenario, an \HII\ region ionizes the
surrounding molecular gas and drives a shock front into the cloud, accumulating a
dense layer of gas between the ionized gas and the shock front. This compressed
gas layer eventually becomes unstable and forms a new OB subgroup. Depending on
the density structure of the surrounding cloud, the compressed gas layer may
trigger star formation in several different ways, ranging from radiation-driven
implosion of pre-existing clouds to a collect-and-compress scenario of the gas layer (see, e.g.,
\citet{Deharveng10}). The latter mechanism would also work for a \CIIr\ region,
but instead of triggering OB stars, it would trigger low-mass stars.  A good
example of  collect-and-collapse triggered star formation is the bright  \HII\ 
region RWC\,79, which shows a dust emission ring surrounding the ionized gas, 
in which \citet{Zavagno06} found two massive fragments that were diametrically
opposite to each other, each containing bright near-infrared sources. They
identified 12 luminous Class I sources toward the most massive of these  dust
fragments. It is often difficult, however, to judge whether the expanding \HII\
region has triggered new star formation or just revealed pre-existing dense
cores in the surrounding cloud. NGC\,2023, however,  looks very similar to
RWC\,79 in the sense that all the young Class I sources are near the periphery
of the \CIIr\ region. Therefore, their formation may well have been triggered by
the expanding \CIIr\ region.

It is clear that when we study the properties of outflows from
young stars, both high spatial resolution and sensitivity are required to
see large-scale structures and small details. With our moderate
spatial resolution, $\sim$ 15\arcsec, we can characterize all the
outflows in the vicinity of NGC\,2023, but we cannot say with certainty
whether it is MIR-63a or b that drives the large southeast-northwest
outflow. The CO(2--1) images by \citet{Luo23} are heavily filtered
by the ALMA interferometer and do not give a clear view of the two
outflows. We assigned the outflow to MIR-63a because \citet{Tobin20} found it  
to be elongated north-south in continuum at 0.87\,mm. If the continuum emission
traces the accretion disk, it would be roughly orthogonal the outflow. It is also 
the more luminous star in the binary system.  Nevertheless, it would be
very useful to  combine the ALMA data  with single-dish data in order
to better determine the morphology and origin of each outflow in the
region in which they are launched.

\section{Summary and conclusions}

The reflection nebula NGC\,2023 is  illuminated by the Herbig Be star
HD\,37903. It is the most massive member of a small young cluster with about 30
PMS stars, a few of which are Class I objects and still accrete heavily. We
therefore expected that they might power molecular outflows. By analyzing
previously published data,  we found four molecular outflows. These four outflows
are all powered by young stars  and are all located in the immediate vicinity of
the reflection nebula. All the outflows have roughly similar ages, suggesting
that the stars driving them were formed about the same time, which supports the
hypothesis that they may have been triggered by the expanding \CIIr\ region.
There are no Class 0 sources in the vicinity of NGC\,2023, although young
protostars like this exist in the large L\,1630 molecular cloud, but they are obviously
unrelated to the young NGC\,2023 cluster.

\begin{acknowledgements}
We would like to thank an anonymous referee for the valuable comments and
suggestions which helped  to improve the manuscript. BM acknowledges the support
of the Department of Atomic Energy, Government of India, under Project
Identification No. RTI 4002. We also thank Dr. W. Varricattu for advice on how
to analyze the UKIDDS images. The paper is based in part on observations made
with the NASA/DLR Stratospheric Observatory for Infrared Astronomy (SOFIA).
SOFIA is jointly operated by the Universities Space Research Association, Inc.
(USRA), under NASA contract NNA17BF53C, and the Deutsches SOFIA Institut (DSI)
under DLR contract 50 OK 0901 to the University of Stuttgart. This work is based
in part on data obtained as part of the UKIRT Infrared Deep Sky Survey.

 \end{acknowledgements}

{}

 \newpage

\begin{appendix}
\section{Position-velocity diagrams}

We plot position-velocity diagrams of CO(3--2) and CO(4--3) for the large southeastern outflow from MIR-63, that is, 
cutting through both the northern and southern red peaks as well as through the outflow from MIR-62.

\begin{figure}[h]
\includegraphics[width=0.48\textwidth]{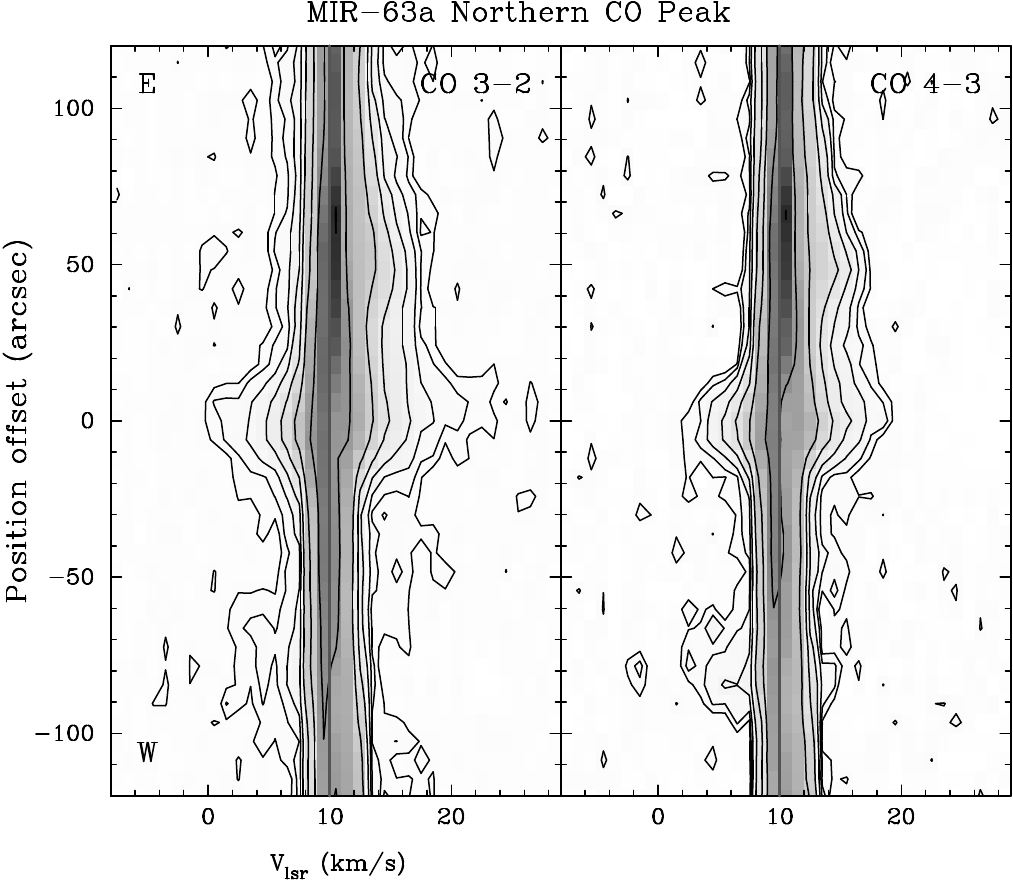}
\caption{Position-velocity diagram plotted  with contours and overlaid with grayscale 
through the northern redshifted peak of the large outflow powered by MIR-63a, i.e., at a PA $\sim$84\degr. 
\label{fig_DNpvplot}}
\end{figure}

\begin{figure}[h]
\includegraphics[width=0.48\textwidth]{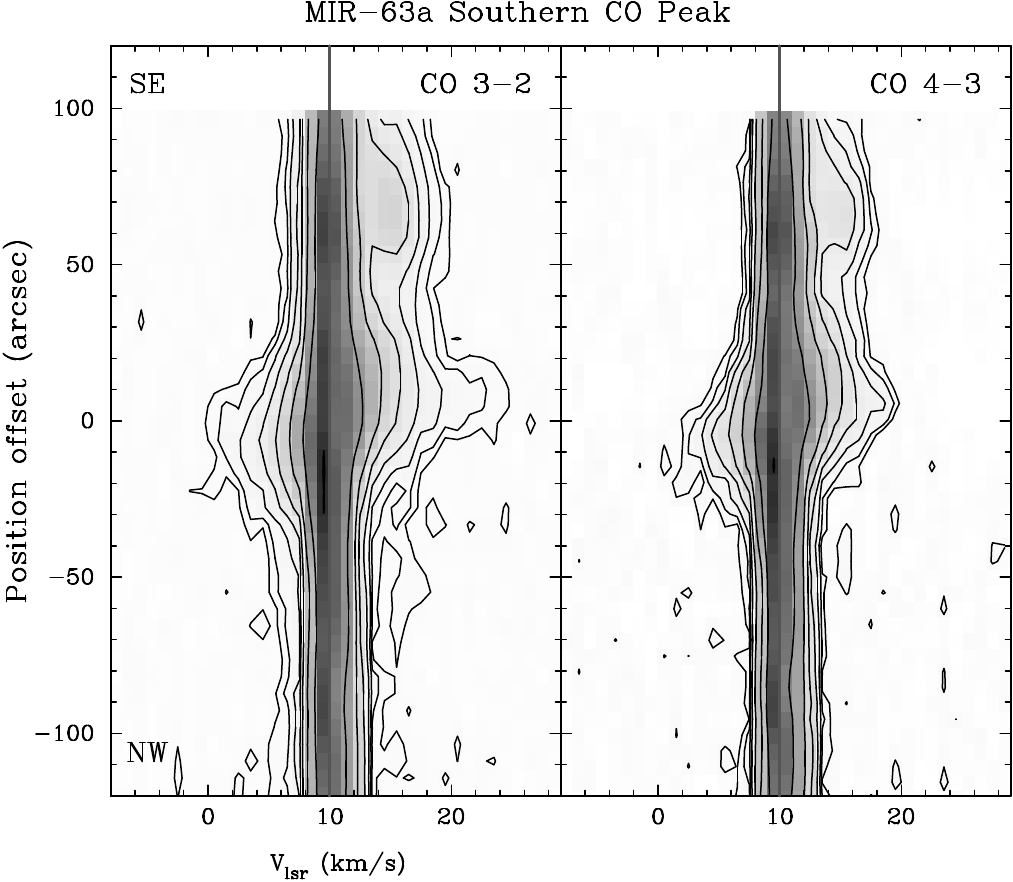}
\caption{Position-velocity diagram plotted  with contours and overlaid with grayscale 
through the southern redshifted peak of the large southeastern outflow powered by MIR-63a, i.e., at a PA $\sim$118\degr, 
\label{fig_DSpvplot}}
\end{figure}

\begin{figure}[h]
\includegraphics[width=0.48\textwidth]{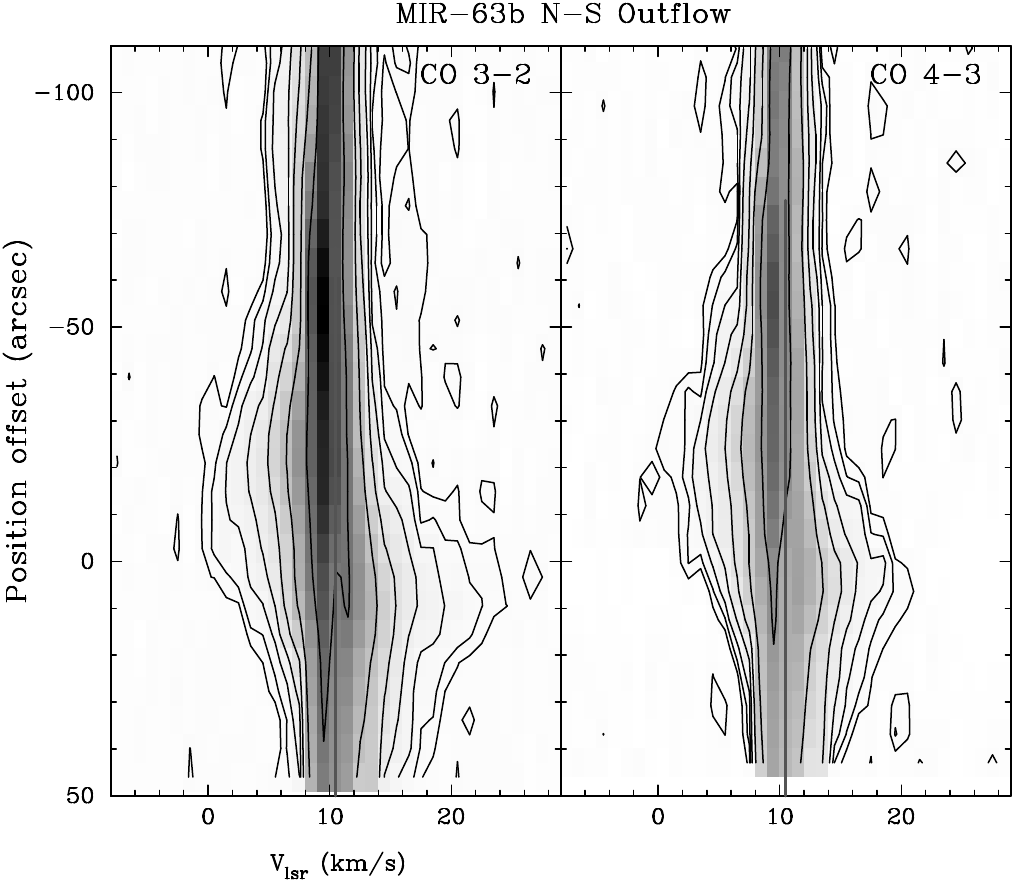}
\caption{Position-velocity diagram plotted  with contours and overlaid with grayscale 
 of the north-south  outflow powered by MIR-63b, i.e., at a PA of -10\degr. 
\label{fig_NSpvplot}}
\end{figure}

\begin{figure}[h]
\includegraphics[width=0.48\textwidth]{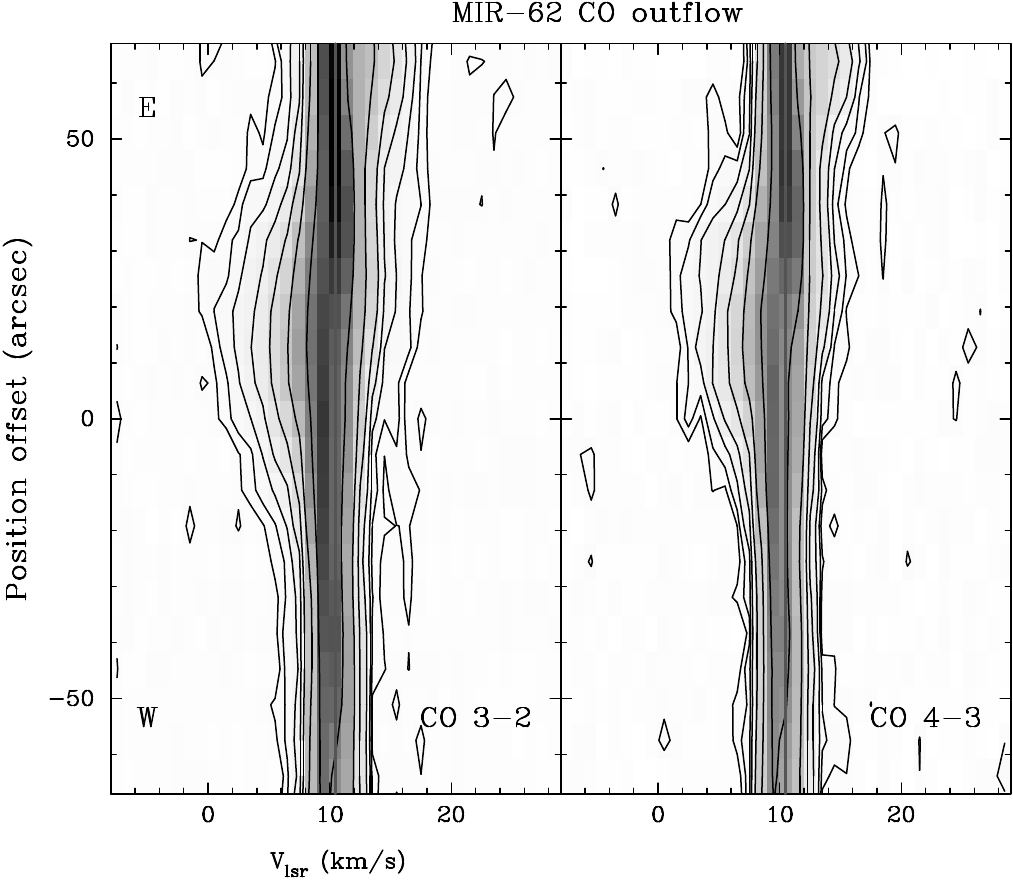}
\caption{Position-velocity plot with contours and overlaid with grayscale 
along the outflow axis of the blueshifted outflow from MIR-62, i.e., at a PA $\sim$110\degr. 
\label{fig_Cpvplot}}
\end{figure}

\end{appendix}

\end{document}